# On the design of an innovative solution for increasing hazardous materials transportation safety

Emil PRICOP

*Graduate student member IEEE*
Petroleum-Gas University of Ploiești
Ploiești, Prahova, România
e-mail: emil.pricop@upg-ploiesti.ro

*Abstract*—Transportation of hazardous materials represent a high risk operation all over the world. Flammable substances such as oil, kerosene, hydrocarbons, ammonium nitrate or toxic products are shipped every day on busy roads by trucks. An innovative solution for increasing hazardous materials transportation safety is presented in this paper. The solution integrates three systems: one mounted on the truck that can alert authorities in case of an accident, one portable system for quick identification of the carried substances and intervention method and a component for real-time road monitoring. The proposed solution is based on RFID card with a special memory structure presented in this paper

## I. INTRODUCTION

Transportation by trucks of hazardous materials, such as petroleum, oil, kerosene, hydrocarbons, liquefied petroleum gas (LPG) and other toxic products poses very high risks to people. In case of an accident a huge number of lives can be threaten if the authorities' intervention is delayed or incorrect. In Romania, it is well known the case of a truck loaded with ammonium nitrate that has exploded in Mihăilești, Romania, killing 18 people and hurting 13 others. This is only one proof that for authorities, especially for fireman and civil protection, is very important to know exactly what substances the truck is carrying and also how to intervene in order to prevent a disaster.

In the last decade many companies has developed various Automatic Crash Notification (ACN) systems. A well-know ACN system is the one built by OnStar company [5] and integrated on some Chevrolet, Cadillac and Buick car models. Some car manufactures such as BMW [1], [7] integrate ACN systems in their new launched models. All those systems detect crashes and alert authorities regarding the accident location. That kind of system is not available in any car or truck and cannot be customized to respond user or authorities requirements.

eCall is a system developed in European Union, having the purpose to give rapid response to people involved in a car accident [4]. This initiative has no component related to the transportation of hazardous materials.

In this paper the author present a road and truck safety solution that can be easily integrated in existing trucks and infrastructures. The developed systems are able to provide authorities with accurate and real-time information regarding the accident location, transported substance and intervention method. The proposed solution has three main components: an integrated system mounted on the truck, a portable system for authorities' usage in case of an intervention and a low-power system that can be integrated in road monitoring infrastructures.

## II. SOLUTION DESIGN AND SPECIFICATION

In this paper we propose an innovative and complex solution having the objective to improve the hazardous materials transportation safety. It has three main components which can be used independently or integrated in a large scale system.

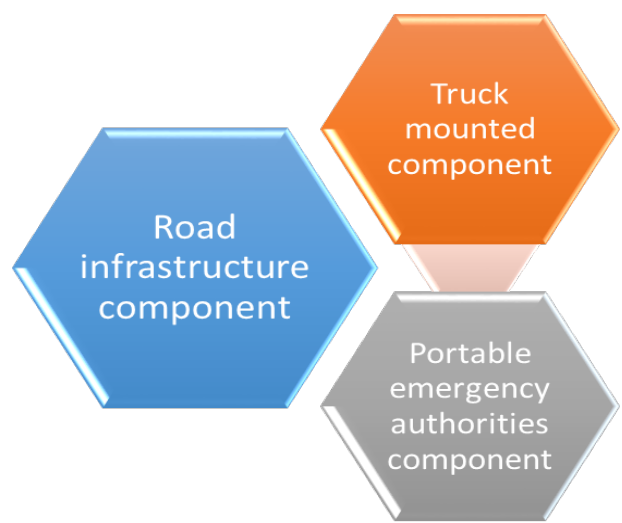

Fig. 1. Proposed solution components

The specifications and purpose of each component of the proposed solution is presented in the following paragraphs. Each component has its own dedicated section in this paper, where design, architecture, integration and functioning will be detailed.

The truck mounted component provide two functions:

- alert the driver when an inappropriate driving manner is detected, for example excessive speeding;
- alert authorities (firemen, ambulance, police) in case of an crash, providing accurate position information along with data regarding transported materials.

This component can be easily extended to function as a real-time monitoring device. It can send position data, speed, acceleration using a GSM connection to the transporter headquarter.

The system will not interfere in any way with the existing electronics in the vehicle. It is completely independent, providing its own sensors, communication interface and power supply. Given this we can assure its flexibility, easy installation and independence from any constraints imposed by car manufacturer or governmental/legal authorities.

The portable emergency component will provide accurate information regarding the substances loaded in the truck. This low-cost, low-power device is based on RFID technology usage. The device integrate a touch-screen that can provide enough information to the authorities regarding the intervention mode.

The road infrastructure component is also based on radio technologies. Long-range RFID readers are mounted from point to point on the road, in order to identify trucks carrying hazardous materials, equipped with RFID tags. Each RFID reader communicate with its neighbors using ZigBee technology. Each 10 readers are allocated to a GSM/GPRS communicator that concentrate data and transmit it to a road monitoring center or authorities involved in traffic safety as police, firemen, etc. This component allows real-time monitoring of road infrastructures regarding transportation of dangerous goods and hazardous substances.

These components can function independently or can be integrated in a large scale system that can monitor and direct interventions in case of crashes or accidents. The integrated system diagram is presented in the figure below:

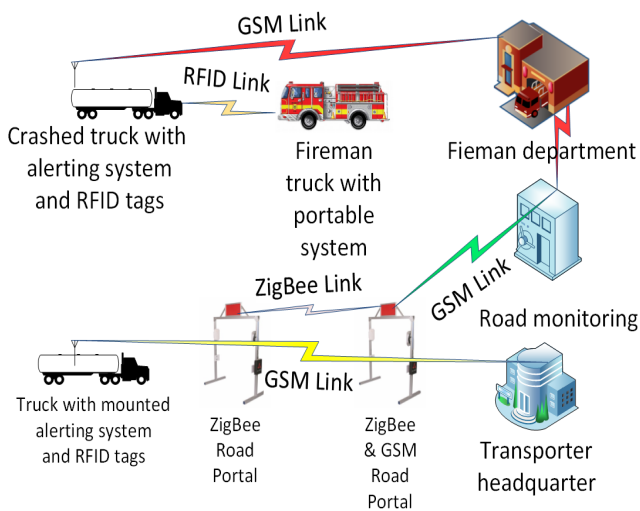

Fig. 2. Integrated system diagram

In Figure 2, there are revealed all the system elements and communication links between them. The truck is the upper side of the image is a crashed one. The system detected the crash and alerted the authorities (fire department, police, ambulance) using the GSM link. The fire department alerts road monitoring institutions in order to block the road and warn the other traffic participants on the accident. All the fire trucks sent to intervene will know exactly what substances are loaded in the crashed truck in the moment of receiving the alert. In a special situation, when the GSM communication between crashed truck alerting system and fire department is not available the forces that intervene and are equipped with the portable system component can read RFID tag on the truck and receive locally details on the transported substances. The RFID reading can be done at a distance between 12-20 meters.

The road infrastructure component is presented in the bottom side of the picture. The truck is equipped with RFID tags and alerting system. On the road from point to point are mounted RFID portals, as needed. To maintain the low cost of the proposed solution only some of the RFID portals will have GSM modems embedded and will function as data collection devices (data aggregators). The other portals, without GSM connection will be equipped with ZigBee transponders that will allow point-to-point communication between two successive portals and then to the GSM enable one in the area. The GSM enabled portal send data to the road monitoring department allowing the authorities to know which truck is transiting the area and what is it transporting.

The central element of each of the three sub-systems presented above is the RFID card that is associated with each truck and the global system database.

III. RFID CARD SPECIFICATIONS

Each truck transporting dangerous good has to be marked with at least one RFID tags in order to be correctly identified. Given the low-cost of the RFID cards and taking into account the experimental results we concluded that it is optimum to use 6 RFID tags on each truck. They should be mounted as follows: one in the front part of the truck, two on the left side, two on the right side and one at the back of the truck, in order to achieve best visibility in movement or in case of an roll-over. The RFID tags are not omnidirectional so there is needed the proposed alignment.

The tags used in this system should be read from a moving vehicle situated at an appreciable distance, so RFID tags on UHF (900 MHz) band should be used. Their reading distance is around 3 meters, but can be extended depending on the reader performances. The reading distance is a critical component especially for the road infrastructure elements of the proposed solution. The tags chosen to be used in the system are re-writable tags with at least 512 Bytes of memory.

Another important aspect is the RFID card memory size and structure. The data stored on the card should provide enough information for the portable system to operate off-line, without any link to the database. In order to achieve this goal the card used has a total memory of 512 Bytes and a special memory structure presented below:

- C_ID – tag identifier is a unique numeric value of 4 bytes;
- T_ID – truck identifier is a unique identifier with length of 8 bytes allocated to each truck that is enrolled in the system;
- T_RN – truck registration number stored as an ASCII value;
- OP_ID – truck operator (transporter) identifier – unique in the database of the proposed system;
- OP_Name – 128 bytes of ASCII data representing transporter organization name in clear text;
- OP_Phone – 13 ASCII characters representing the phone number of the truck operator. From this phone number authorities can gain valuable information regarding the substance transported.
- S_id – unique code defining each transported substance. These codes are defined by the system developer and are global variables. For example methane can have code 2C in hexadecimal. Field length is 8 bytes.
- Comp-id – memory zone allowing to store all the components codes for transported substances in case of a mixture. The field length is 40 bytes.
- Ign_p – ignition point for the transported substance, stored as ASCII character, maximum length 5 bytes;
- Sig_temp – self-ignition temperature stored as ASCII text on 5 bytes;
- Exm_id – extinguishing materials identifiers. Each material has its own unique id number. The field length is 60 bytes.
- B_Pnt – the boiling point of the transported material. It will be stored in ASCII format on 4 bytes;
- M_Pnt – the melting point of the transported material. It is stored on 4 bytes as ASCII text;
- S_Dens – substance density is represented on 8 bytes as ASCII text;
- Tox_V - this field of 2 bytes show whether the carried substance is toxic or not;
- Kemler_no – the Kemler number is stored as an ASCII text value on maximum 8 bytes;
- Onu_no – the ONU number is stored as an ASCII text value on maximul 8 bytes;
- Et_id - field of 20 bytes storing codes for each tag defined by law to be attached to the truck;
- User-def – 40 bytes of user defined data;
- Ecc – is the hash of the previous data, automatically generated using a cyclic-redundancy check algorithm. This field has a length of 8 bytes.

The RFID card defined above permits to create a new generation of electronic records that can completely identify a truck transporting a hazardous or dangerous material. The data stored in the RFID card's memory comply with actual transportation laws and regulations in Romania [6] so we can assume it can replace the paper documents used at a large scale now.

The RFID tag is a local component of the system. It allows rapid identification of the truck, loaded substances and gives accurate information regarding substance properties and intervention ways for authorities in case of an emergency.

We proposed the usage of re-writable RFID tags in order to keep the costs associated with system exploitation low. The RFID card containing all the needed information is issued at the transporter headquarter when the truck is loaded.

IV. TRUCK MOUNTED COMPONENT DESIGN

Systems that are able to alert authorities when a crash or car accident happened have been developed by various companies and research institutions. The system presented in the paper is innovative because it can transmit data regarding the transported goods or substances. In this way the authorities are prevented about the risks posed by the crash and they can intervene with corresponding means, various neutralizing substances, etc. Another component not present in existing similar systems is alerting the driver when a dangerous manner is detected. The proposed system measure speed and acceleration and plays a corresponding pre-recorded message. It is also possible to use a display to show graphical warnings around the car dashboard.

The block diagram of the system is presented in Figure 3.

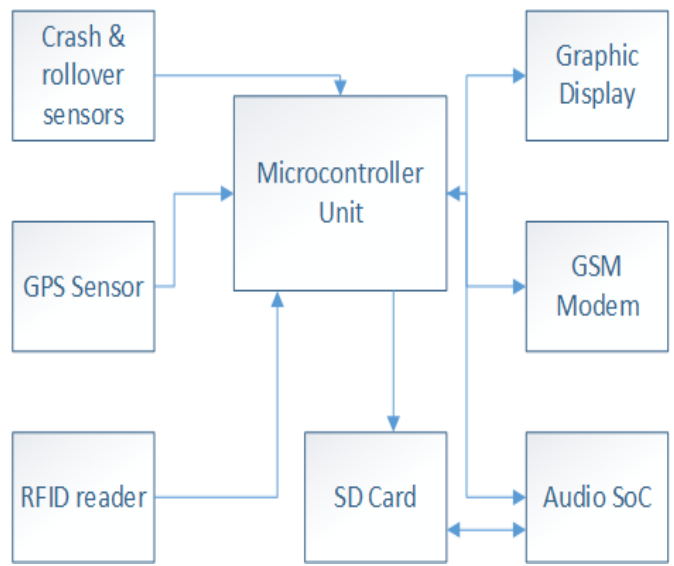

Fig. 3. Block diagram of the truck mounted system

The system uses crash & rollover sensors as presented in [1], [2], [3]. Basically the crash sensor uses two or more accelerometers to detect sudden acceleration changes and pressure sensors mounted in specific places of the car. Those crash detectors are similar to the ones implemented in the airbag triggering system and are not the objective of this paper.

The GPS sensor is a standard SIRF Star III module. This is a System-on-a-chip (SoC) module integrating a powerful DSP, a MCU unit and having 20 channels. The external antenna is mounted in a non-shielded zone of the vehicle.

The RFID reader has the role to read one of the RFID tags mounted on the truck. In case of a crash it will retrieve all the necessary details to provide authorities with accurate and complete information regarding the substance transported.

The SD Card is used as a recording device for storing GPS and crash sensor data. Also the same memory card can be used by the audio SoC as a source for MP3 files that will be played when an inappropriate driving manner is detected.

As shown in Figure 3 there are three main output devices. The graphical display and audio SoC are used for local alerting the driver. The GSM modem is a standard module with GSM (voice and SMS) or GPRS (data transmission) capabilities that is used to alert authorities in case of a crash.

The device can be viewed as a "black-box" device, as the ones mounted on the planes, but with an extended function of alerting authorities in case of an accident.

When a crash is detected the crash and rollout sensor transmits a signal to the microcontroller. This signal will trigger the alerting state. The MCU will command the RFID reader to retrieve data from the RFID tag in its nearby and send all the data together with an alarm string to the authorities dispatch.

In normal working mode the device will continuously measure driving speed, direction and acceleration by using GPS sensor and accelerometers in the crash sensor. When the MCU detects speeding over a pre-defined value (e.g. 70 km/h) or a rapid change in acceleration, it will trigger the audio SoC to play a prerecorded audio message and will display a corresponding visible sign on the graphical display.

The experimental model for demonstrating the concept of the proposed system is based on Arduino development board and specific peripherals (shields). Arduino [8] is a open-source development platform based on ATMEL ATMega 328 [9] microcontroller with all the necessary peripherals embedded on the same circuit board. For our prototype we have used an RFID reader shield developed by Sparkfun [14], featuring an ID Innovation ID-12 [12] reader chip. As a rollout sensor we have used an ADXL345 accelerometer [10]. The GPS receiver used was mounted on a shield produced by Sparkfun and is based on GlobalSat EM-406A GPS receiver [14]. Data transmission has been done using a GSM shield based on QuecTel M10 GSM module [13]. As output devices the prototype integrates an MP3 player trigger shield powered by Cypress PSoC CY8C29466-24SXI [11] microcontroller and VS1063 audio codec and a standard graphical LCD with 4 lines x 128 columns. All the peripherals were integrated and programmed using standard libraries for Arduino [8].

The tests were done in a laboratory of Petroleum-Gas University of Ploiesti, Romania. The microcontroller was programmed to send the content of the RFID tag read to a server in the laboratory via GPRS connection, when there is detected a suddenly acceleration variation. Simultaneously the prerecorded message was played and on the LCD was displayed the following text: "ALERT ACCIDENT". The tests concluded that the design of the solution is practical and that it can be used on a large scale.

## V. PORTABLE EMERGENCY COMPONENT

The portable emergency component is designed to display information regarding the truck and carried products at the intervention place without a data-link back to the dispatch. This component is very useful in case of an accident that took place in an area without GSM coverage or when no information is provided via the dispatch. It can be also used by police or road management authorities in order to obtain information on transported goods.

The proposed device include a long range RFID reader, which allows getting RFID tag data from a distance of at least 12 meter, an SD Card for storing a local information database and a touchscreen display that provide information in a friendly user-interface. The block diagram of the system is presented in Figure 4.

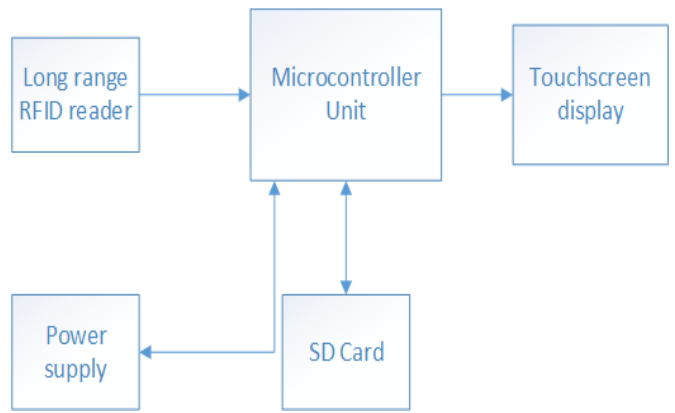

Fig. 4. Block diagram of the portable emergency component

The device functioning is not so complicated. The long range RFID reader get data from a tag placed on the truck. Data is transmitted to the microcontroller unit (MCU) that start processing it according to the RFID card details described in Section III of this paper. The MCU parses the data received in order to get value for each field defined. When a code is found, for example $S\_id$, the MCU searches in a database stored on the SD card for the corresponding substance name.

Taking into account that the device is portable we assume it has constraints regarding power consumption, dimensions and weight. The long range RFID reader and the touchpad display are high energy demanding components, but they are essential for the correct system functioning. A high capacity rechargeable battery with a minimum capacity of 3000 mAh should be used. The power supply block will also include a recharging circuit that allows recharging the battery from the cars cigarette lighter (12V).

The experimental model for this device is also built using Arduino development board. The RFID reader can be ID Innovation ID-12 or any UHF reader with serial communication. The touchscreen shield uLCD-43-PT-AR [Z] used in tests was developed by 4D Systems. It features a 4.3" color LCD touchscreen with advanced graphics functions and serial communication.

As shown in Figure 4 the portable device does not include any communication module. We have chosen this design taking into account the price of the GSM modules and the power consumption and weight constraints. Given these considerations the system database has to be a local one, stored on the SD Card. implement a real database on a MCU is a difficult task due to the limited resources available, so we have chosen to use text files in a special folders structure. Each field defined for the RFID tag memory structure represents a folder that include text files named with the possible codes for the respective field. Each text file contains specific information, basically ASCII text that should be displayed to the user. This search algorithm is not optimal and a new research direction identified when working on this paper is the implementation of a high-performance relational database management system on a MCU with storage on a SD card.

## VI. ROAD INFRASTRUCTURE COMPONENT

The road infrastructure component represents the base element of an integrated road monitoring system. There are some attempts to build intelligent road infrastructures [16], but most of them focus on giving the driver information on traffic conditions, not on monitoring the dangerous materials transported.

The proposed system relies on the usage of a portal with a long-range RFID reader and 4 UHF antennas mounted in selected points on the road in order to detect trucks and read at least one of the RFID tags mounted on them. The portal include 4 antennas because of the directional properties of RFID electromagnetic waves. The antennas will be mounted symmetrically on each side of the road on two pillars.

When a truck passes through the portal the RFID reader detects and reads data at least one of the tags placed on the truck. The data received is sent to the dispatch of the road monitoring institution via a secure GSM channel. Optionally data can be logged locally on an SD card. Since the data is already sent via GSM connection to the monitoring dispatch, the SD card is only used as a backup local storage device. In the further development of the system it should be replaced with a storage device with fire-resisting capabilities.

In order to send data to the authorities dispatch a GSM/GPRS connection is used. GSM equipment and data services are not cheap so we proposed a combined solution: GSM enabled portals will be mounted at an appreciable distance. If needed between them can be installed ZigBee or XBee enabled portals, creating a point-to-point network connected to a GSM enabled portal. By using mixed XBee and GSM portals a true road monitoring network infrastructure is built.

. The microcontroller unit receives data from the RFID reader circuit and commands the GSM modem to forward it to the dispatch. As shown in figure 5 the device can work as a relay for the XBee-only (figure 6) road portals, collecting data via the XBee embedded module and forwarding it to the dispatch using the GSM connection. Since the device integrates an SD Card reader all the data can be stored locally for logging purposes.

The GSM-enabled road portal block diagram is presented in the figure below

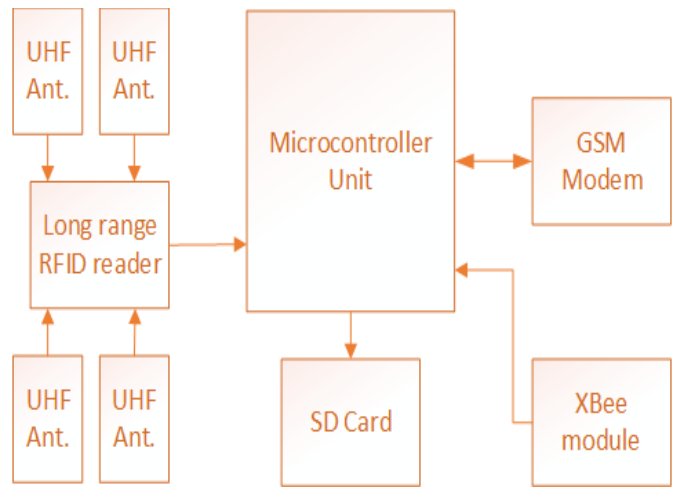

Fig. 5. Block diagram of the road infrastructure component with data collecting capabilites

For testing purposes we have integrated a long-range UHF RFID reader and 4 UHF antennas (900 MHz) with an Arduino board and a standard GSM shield. The results were satisfactory, the system being able to read tags from 12 meters even in moving. In order to implement the solution the Arduino board should be replace with a custom board using the same ATMEL MCU and integrating only absolutely necessary peripherals and a cheap GSM module.

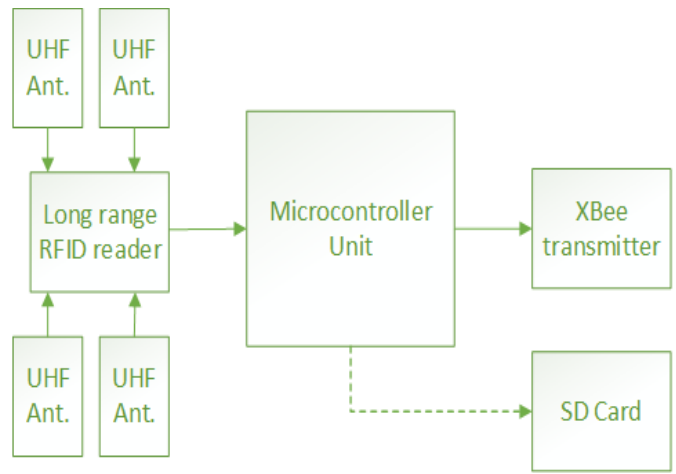

Fig. 6. Block diagram of the road infrastructure component with ZigBee transmitter

In the figure above is presented the block diagram of the road infrastructure component with integrated XBee transmitter. The RFID components are similar to the ones integrated in the device presented in figure 5. The SD card block is optional and can be removed in order to keep the price low.

This device collect data from the truck with embedded tags. The microcontroller, ATMEL ATMega 328, decodes data from the RFID reader and transmit it to a GSM-enabled portal.

The road monitoring dispatch collect data from all the GSM enabled portals in real-time. The authorized users can generate reports or track in real-time the movement of trucks. The database structure and web application design and functioning will be presented and analyzed in a future work, since this paper focus mainly on the RFID tag memory structure and hardware components.

## VII. CONCLUSION

The solution proposed in this paper integrates three complex stand-alone systems based on a special memory structure of an RFID card in order to increase the safety of hazardous materials transportation. Each device presented herein can be used as a stand-alone equipment for signaling a crash to the dispatch, quickly and exactly identifying dangerous goods transported using a truck or monitoring in real time the existent road infrastructure.

The central component of the designed system is an RFID tag with a specially design of its memory structure as described in Section III.

The whole solution was developed using Arduino development board and compatible peripherals. The developed systems were laboratory tested, confirming in this way the design and desired functionality. The Arduino development board should be replaced with a compatible microcontroller (ATMEL ATMega 328 MCU) and the peripherals should be integrated on the same board in order to reduce costs and size for mass-production and large scale implementation of the proposed solution.

The system can be further developed in order to implement an integrated road monitoring system based on intelligent infrastructures (sensor networks, "talking cars", RFID enabled devices, etc.). The beneficiary of the system could be governmental authorities along with transporters who carry hazardous goods and substances.

## REFERENCES

[1] Ertlmeier, R.; Spannaus, P., "Expanding design process of the Airbag Control Unit ACU - Connection of Active and Passive Safety by using vehicles dynamics for rollover and side crash detection," Intelligent Solutions in Embedded Systems, 2008 International Workshop on , pp.1,9, 10-11 July 2008

[2] Mahmud, S.M.; Alrabady, A.I., "A new decision making algorithm for airbag control," Vehicular Technology, IEEE Transactions on , vol.44, no.3, pp.690,697, Aug 1995

[3] Rauscher, S., Messner, G., Baur P., et al., Enhanced automatic collision notification system-improved rescue care due to injury prediction-first field experience." in Proceedings of the 21st International Technical Conference on the Enhanced Safety of Vehicles (ESV), Stuttgart, Germany. 2009.

[4] ***, ECall Website – European Council -http://ec.europa.eu/digital-agenda/ecall-time-saved-lives-saved

[5] ***, OnStar Company Website – www.onstar.com

[6] ***, ISU Buzau – Regulations regarding the transportation of hazardous substances http://www.isubuzau.ro/legislatie_insp_prevenire/Documentar%20substante%20periculoase.pdf

[7] *** - http://www.psfk.com/2011/07/bmw-puts-your-safety-first-with-new-automoatic-crash-notification-technology.html

[8] ***, Arduino Development Board documentation – http://www.arduino.cc

[9] ***, ATMEL ATMega 328P Microcontroller Datasheet - ATMEL Company - http://www.atmel.com/Images/doc8161.pdf

[10] ***, ADXL345 datasheet - Analog Devices Corp. website http://www.analog.com/static/imported-files/data_sheets/ADXL345.pdf

[11] ***, Cypress Company, website – www.cypress.com

[12] ***, ID Innovations Incorporated website – www.idinnovations.com

[13] ***, Quectel M10 GSM Module datasheet - http://satronel.com/product/Quectel/M10

[14] ***, Sparkfun Electronics Company website – www.sparkfun.com

[15] ***, uLC-43-PT-AR documentation, 4D Systems Corp. website - http://www.4dsystems.com.au/product/3/21/Arduino_Display_Modules_and_Shields/uLCD_43_PT_AR/

[16] ***, Hessian Ministry of Economy, Transport, Urban and Regional Development, Wiesbaden, Germany – Ambient Mobility Brochure 2010, http://www.hessen-it.de/mm/brochure_ambient_mobility.pdf